\documentclass[10pt,conference]{IEEEtran}
\PassOptionsToPackage{hyphens}{url}
\usepackage{hyperref}
\usepackage{amsmath,amsfonts}
\usepackage{algorithmic}
\usepackage{array}
\usepackage[caption=false,font=normalsize,labelfont=sf,textfont=sf]{subfig}
\usepackage{textcomp}
\usepackage{stfloats}
\usepackage{url}
\usepackage{verbatim}
\usepackage{graphicx}
\usepackage{balance}
\usepackage{pifont}
\usepackage[utf8]{inputenc}
\usepackage{booktabs}
\usepackage{xcolor}
\usepackage[most]{tcolorbox}
\usepackage{fontawesome5}
\usepackage{pgfplots}
\pgfplotsset{compat=1.18}
\usepackage{tabularx}
\usepackage{colortbl}
\usepackage{tikz}
\usepackage{cite}
\usepackage{todonotes}
\usepackage{subcaption}
\def\BibTeX{{\rm B\kern-.05em{\sc i\kern-.025em b}\kern-.08em
    T\kern-.1667em\lower.7ex\hbox{E}\kern-.125emX}}
\definecolor{quotepurple}{RGB}{87, 47, 150}
\definecolor{takeawaybg}{RGB}{222, 213, 234}
\definecolor{lakerspurple}{RGB}{85, 37, 130}
\definecolor{lakersgold}{RGB}{253, 185, 39}
\definecolor{accent}{HTML}{2A6DB0}
\definecolor{ro}{RGB}{180, 90, 0}

\newcommand{\pquote}[1]{\textit{\textcolor{quotepurple}{#1}}}

\newcommand{\implic}[1]{%
  \tikz[baseline={([yshift=-0.5ex]current bounding box.center)}]{%
    \node[fill=accent, rounded corners=2.2pt, minimum size=2.1ex,
          inner sep=0pt, text=white, font=\footnotesize]{\faLightbulb};%
  }\hspace{0.4em}\textcolor{accent}{\textbf{#1}}%
}

\definecolor{conditionalcolor}{HTML}{F3E7C7}
\newcommand{\conditionalbox}{%
  \raisebox{0.1ex}{\color{conditionalcolor}\rule{0.9em}{0.9em}}%
}

\begin{document}
\title{How Do Practitioners Build SE Agents? \\ Insights from a Mixed-Methods Study}

\author{
\IEEEauthorblockN{%
Yunbo Lyu\textsuperscript{1}, 
David Williams\textsuperscript{2}, 
Jieke Shi\textsuperscript{1},
Zhensu Sun\textsuperscript{1,*}, 
Chao Peng\textsuperscript{3}, 
Zhou Yang\textsuperscript{4,5},
Federica Sarro\textsuperscript{2}, 
David Lo\textsuperscript{1}%
}
\IEEEauthorblockA{%
\textsuperscript{1}Singapore Management University, Singapore\quad
\textsuperscript{2}University College London, UK\\
\textsuperscript{3}Tencent, China\quad
\textsuperscript{4}University of Alberta, Canada\quad
\textsuperscript{5}Alberta Machine Intelligence Institute, Canada\\[2pt]
\{yunbolyu, jiekeshi, zssun, davidlo\}@smu.edu.sg,\quad
\{david.williams.22, f.sarro\}@ucl.ac.uk,\\
chao.peng@acm.org,\quad zy25@ualberta.ca\quad \textsuperscript{*}Corresponding author
}
}

\maketitle

\begin{abstract}
The rise of Software Engineering (SE) agents, i.e., LLM-based agents that can understand large codebases and carry out engineering tasks with limited human intervention, has been marked by rapid advance and adoption, but little is known about how developers build these systems in practice: existing studies mine repositories or examine deployment, but few investigate how SE agents are constructed. 
Through semi-structured interviews with 20 practitioners from 12 organizations and an online survey of 80 practitioners, this paper is the first to study how SE processes are changing in the development of SE agents and what challenges developers face. 
We find that as implementation becomes cheaper, bottlenecks shift rather than disappear: long-standing work in requirements, coordination, and deployment becomes more visible, while reviewing generated code and evaluating agent behavior become new and increasingly central forms of work.
We characterize a seven-stage workflow and five process shifts, including a move toward evaluation-driven development, in which evaluation is increasingly defined early and steers iteration, and the emergence of specifications as first-class artifacts that teams test and version alongside code.
We further identify six challenges that teams face, together with 12 corresponding practices they use or propose to address them, including unreliable evaluation signals, comprehension debt as code outpaces understanding, and behavioral changes introduced by provider-side model updates.

\end{abstract}

\begin{IEEEkeywords}
Software Engineering, Agents, AI4SE, AI coding assistants, SE agents, AI, Software Process.
\end{IEEEkeywords}

\section{Introduction}
\label{sec:intro}

Software Engineering (SE) is being reshaped by agents built on Large Language Models (LLMs)~\cite{yang2024swe}, which we refer to as {\it SE agents}. Exemplified by tools such as Claude Code~\cite{anthropic-claude-code} and Codex~\cite{openai-codex}, these agents can carry out complex SE tasks, such as feature development, debugging, and code review, with minimal human oversight. Their adoption is already substantial: Anthropic reported that Claude Code authored more than 80\% of the code merged into its own codebase as of May 2026~\cite{anthropic-whenaibuildsitself}, while Cursor, one of the earliest companies to commercialize SE agents, surpassed \$2 billion in annualized revenue within four years~\cite{techcrunch-cursor}. These trends suggest that SE agents are becoming routine actors in software development rather than experimental tools.

Meanwhile, it is easy to overlook that SE agents are themselves software systems, yet ones whose behavior depends not only on conventional code, but also emerges from the underlying model, prompts, tools, context, and execution environment. As software, they should in principle be amenable to an effective development process; however, because they differ fundamentally from conventional software, that process cannot simply be carried over. This difference is most apparent in debugging: when an SE agent misbehaves, there may be no single faulty statement to repair, since the failure can originate in the model, the harness,\footnote{We use \emph{harness} to refer to the system layer that wraps the model and manages its execution, and treat \emph{scaffold} as a synonym.} a tool interface, or their interaction. The familiar edit-run-debug loop no longer offers a concrete point of intervention. More broadly, when behavior is determined by a model rather than by the code one wrote, the traditional anchors of software development shift, raising the question of how SE agents should be specified, developed, evaluated, deployed, and maintained.

However, little is known about how these processes actually unfold in practice. Existing empirical studies primarily mine repositories, issues, commits, and developer discussions to examine how agent-based software is built~\cite{wang2025empirical,asgari2025challenges,hasan2026empirical}. Interview studies of agentic systems have focused mainly on their production deployment rather than their development~\cite{pan2025measuring}, while studies of AI-based software largely predate the agentic shift and center on integrating machine learning or LLM components~\cite{nahar2022collaboration,wan2021how,nahar2025beyond,parnin2025building}. To our knowledge, no interview study has investigated how practitioners build SE agents.

Our study fills this gap by combining semi-structured interviews with 20 practitioners from 12 organizations with an online survey of \textcolor{black}{80} practitioners with hands-on experience in building SE agents. The organizations ranged from startups to mid-sized companies and large technology companies at the forefront of SE agent development. 
Notably, many of these practitioners use coding agents to build SE agents themselves, occupying the dual role of builders and early adopters.
We conducted each interview following a semi-structured guide and analyzed the transcripts using thematic analysis and a hybrid card-sorting approach, deriving a seven-stage agent-building workflow, five process shifts, six challenges, and 12 associated practices. We then performed member checking~\cite{birt2016member}, i.e., sharing findings back with interviewees to confirm the accuracy of our interpretations. 
Following this, we surveyed a wider practitioner population via an online questionnaire, in which most findings received broad overall agreement, thereby corroborating the interview results.

Our results show that building SE agents does not eliminate the traditional development process so much as reorganize it. 
We characterize a recurring seven-stage loop: requirements, evaluation, data, system construction, testing and deployment, human feedback, and adaptive maintenance. Within it, construction follows a cheapest-first model strategy supported by a harness layer shared across model choices, and evaluation recurs throughout the loop. 
We further identify five process shifts. 
Most participants also used SE agents to build SE agents, moving implementation to a higher level of abstraction and reducing the effort required to produce code (S1). 
As coding effort shrinks, effort is both \emph{unmasked} and \emph{created}: long-standing work in requirements, coordination, and deployment becomes more visible, while reviewing generated code and evaluating agent behavior create new and increasingly central work (S2). 
Practice is moving toward \emph{evaluation-driven development} (EDD), in which evaluation is increasingly defined early and steers iteration rather than serving as a final check (S3).
Cheaper runnable artifacts and the combination of model research with production engineering shrink role boundaries (S4). 
Finally, prompts, skills, context definitions, and scaffold behavior become first-class artifacts that teams engineer, test, and version alongside code (S5). 
Our survey supports the overall workflow (91\% agreement) and all five process shifts (71--95\%).

These shifts bring new challenges. 
Evaluation can lose its trustworthy signal: the tests meant to judge an agent become the oracle simply because they exist, extending and inverting the classic test-oracle problem. 
Comprehension debt accumulates as agent-generated code enters the system faster than developers can understand and review it, motivating an emerging alternative we call \emph{regenerative software}: preserving the specifications, tests, and infrastructure needed to reconstruct and validate software on demand.
Finally, the ground shifts: under what we call the \emph{``change nothing, change everything''} effect, provider-side model updates can alter an agent's planning, tool use, and behavior even when the team changes none of its code, prompts, or tools.
The remaining challenges concern safety trade-offs, inaccessible tacit knowledge, and inadequate productivity metrics.
For each challenge, we distill practices that teams use or propose, and our survey provides broader support for most of these practices.

\section{Background and Related Work}
\label{sec:related}

\subsection{Software Engineering Agents}
\label{subsec:ai-coding-assistants}

SE agents are LLM-based systems that combine language models with tools, memory, planning, and environmental feedback to perform SE tasks~\cite{liu2026llmagentsurvey}.
 Unlike earlier coding assistants that mainly generated code completions or isolated snippets, SE agents can inspect repositories, invoke tools and commands, observe execution results, and iteratively decide how to proceed~\cite{yang2024swe,xia2024agentless,wang2024openhands,williams2026pomona}.
 Existing agents have been developed for repository-level issue resolution, program repair, testing, debugging, and software maintenance~\cite{jimenez2024swe,zhang2024autocoderover, gong2025ga4gcgreeneragentgreener,wang2025agents,jin2024llms,lyu2026agentszz}.
Their autonomy varies: some follow predefined workflows, while others dynamically plan and select actions based on feedback from the development environment~\cite{xia2024agentless,bouzenia2025understanding}. 
 Systems such as Claude Code and Codex illustrate this shift toward end-to-end agentic development.  
 In this study, we use \emph{SE agents} to refer to both general-purpose coding agents and agents developed for specific SE activities, such as requirements engineering, testing, optimization, security, and operations.

\subsection{Software Engineering Process}
\label{subsec:se_process}

An SE process comprises interrelated activities, methods, tools, and resources for developing, operating, and maintaining software systems~\cite{washizaki2025swebok}. These activities span the software life cycle, from conception to retirement~\cite{iso24748}, while software life cycle models (SLCMs) define how they are organized, sequenced, and iterated~\cite{washizaki2025swebok}.
SLCMs have evolved by reorganizing when and how engineering activities are performed~\cite{ruparelia2010software}. Waterfall structures requirements, design, implementation, testing, deployment, and maintenance sequentially~\cite{royce1987managing,pressman2005software}, but can be costly under changing requirements and late feedback~\cite{pressman2005software}. The spiral and V-model introduced iteration, risk management, and earlier verification and validation~\cite{boehm1988spiral,forsberg1991relationship}. Agile later emphasized responsiveness, frequent delivery, and customer collaboration~\cite{manifesto2001manifesto}, as reflected in Scrum~\cite{schwaber2001agile} and Extreme Programming~\cite{beck2000extreme}, including test-driven development~\cite{beck2003test}. DevOps extended continuous feedback into operations~\cite{iso32675}, followed by MLOps~\cite{kreuzberger2023machine}, LLMOps~\cite{diaz2024large}, AgentOps~\cite{dong2024agentops}, and evaluation-driven development and operations~\cite{xia2024evaluation}.
ML-enabled development further interleaves data collection, labeling, training, evaluation, and monitoring with conventional engineering~\cite{amershi2019software,wan2021how}. 
SE agents introduce another shift: developers increasingly build SE agents while using coding agents in their own work~\cite{huang2025professional}. 
How this dual role reshapes software engineering remains underexplored.

\subsection{Challenges and Solutions in Building LLM-based Software}

The most relevant study to our work is by Nahar et al.\cite{nahar2025beyond}, who identify 19 emerging solutions for integrating LLMs into software through a mixed-methods study at Microsoft. Another work by the same authors\cite{nahar2023metasummary} synthesizes common challenges across AI development studies. Parnin et al.\cite{parnin2025building} describe the architecture and construction of AI-enabled development tools, while Wan et al.\cite{wan2021how} examine how ML development differs from traditional software engineering. Several emerging concepts also relate to our topic, including LLMOps~\cite{diaz2024large} and AgentOps~\cite{dong2024agentops}, which focus on operationalizing LLM-based and agentic systems. 
Amershi et al.\cite{amershi2019software} provide early insights into AI engineering practices at Microsoft, and Steidl et al.\cite{steidl2023pipeline} and Lau et al.\cite{laudesign} discuss AI development pipelines and AI product design principles. However, existing empirical work largely mines development artifacts or studies production deployment, while interview studies of AI-based software predate the agentic shift\cite{wang2025empirical,asgari2025challenges,hasan2026empirical,pan2025measuring,nahar2022collaboration}. 
To our knowledge, no interview study has examined practitioners building SE agents.

\section{Research Design}
\label{sec:research-design}

\begin{figure*}[t]
  \centering
  \includegraphics[width=0.8\textwidth]{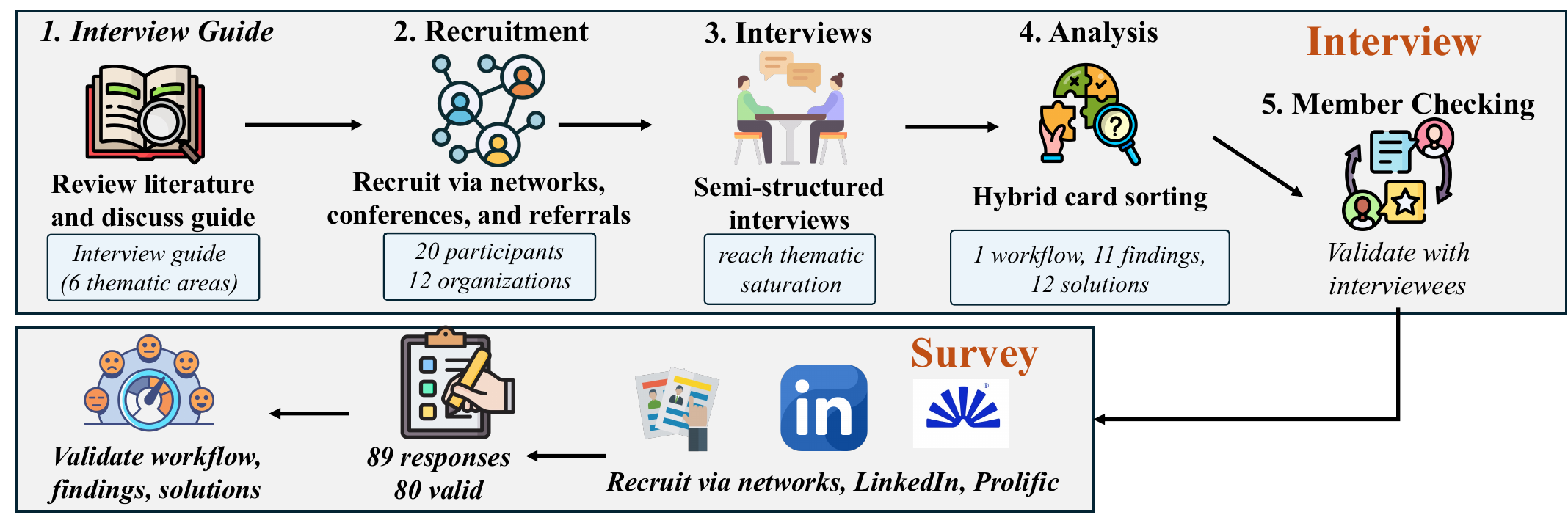}
  \caption{Overview of our research method. 
  We adopt an exploratory sequential mixed-methods design. 
  The interview study proceeds from scoping and interview-guide design, through recruitment, semi-structured interviews, and hybrid card-sorting analysis, to member checking with participants. 
  The resulting findings are then validated through a survey of a broader population of practitioners recruited via personal networks, LinkedIn, and Prolific.}
  \label{fig:method}
\end{figure*}

Fig.~\ref{fig:method} shows the overview of our methodology. 
In this study, we adopt an exploratory sequential mixed-methods design~\cite{creswell2003advanced}, following prior studies~\cite{wan2021how,nahar2025beyond,al2021app, HannaTOSEM}.
We first conduct qualitative interviews and use the findings to develop a subsequent quantitative survey.
Both our interview and survey were \textbf{approved by an Institutional Review Board} (details omitted due to blind review).

\subsection{Research questions}
\label{subsec:research_questions}

As AI capabilities have evolved from traditional machine learning to LLM-based and agentic systems, the practices for building SE agents have undergone substantial transformation.
We seek to investigate how practitioners develop SE agents and what challenges and corresponding practices emerge from their development and use.
Based on these motivations, we formulate two research questions (RQs):

\emph{\textbf{RQ1 (Process). How are SE agents developed in practice, and how are software engineering processes changing?}}
This question investigates (1) the recurring workflow through which practitioners develop SE agents and (2) the broader process changes associated with their development.

\emph{\textbf{RQ2 (Challenges \& Practices). What challenges arise when developing SE agents, and what practices can effectively address them?}}
This question examines the challenges practitioners encounter during SE agent development and the practices used or proposed to address them.

\subsection{Interviews}
\label{subsec:interview}

We conducted a qualitative interview study to explore how developers build SE agents. Following prior qualitative studies~\cite{wan2021how,nahar2022collaboration,HannaTOSEM}, the study proceeded in four stages: (1) we developed the interview guide based on an initial review of related literature; (2) we conducted semi-structured interviews and iteratively reviewed emerging findings until thematic saturation was reached; (3) we finalized the analysis; and (4) we conducted member checking with participants to assess the accuracy of our interpretations.

\subsubsection{Scoping and Interview Guide}
To scope the study and inform the interviews, we conducted a targeted review of prior work on the processes and challenges of building AI-based software systems, particularly AI agent systems (Section~\ref{sec:related}). 
Following prior qualitative research~\cite{nahar2022collaboration}, we searched Google Scholar, the ACM Digital Library, and IEEE Xplore for relevant papers, used Google Search to identify public industry reports, and manually reviewed recent work from leading SE venues.
The first author reviewed the identified sources and, drawing on 16 selected papers and industry reports~\cite{nahar2025product,nahar2022collaboration,nahar2023metasummary,nahar2025beyond,mao2025whatscode,huang2025professional,dong2025survey,anthropic2026evals,huang2025aiwork,matos2025aidlc,wan2021how,amershi2019software,chen2025empirical,asgari2025challenges,wang2025empirical,liang2024large}, drafted the interview guide.
Five experienced SE researchers refined it through two rounds of discussion. 
We then conducted a pilot interview and revised the guide based on the participant’s feedback. 
The final guide was completed in January 2026.

\subsubsection{Semi-structured Interviews}
Following prior studies~\cite{nahar2022collaboration,wan2021how}, we conducted semi-structured interviews with 20 participants from 12 organizations between February and May 2026.
The interview guide covered six areas: (i) participant background and role; (ii) organizational context and team communication; (iii) development processes and workflows; (iv) challenges and practices in building SE agents; (v) developer skillsets; and (vi) open-ended questions.
The complete guide and follow-up probes are available in our replication package~\cite{repo}.
Each interview lasted approximately 60–75 minutes. 
All participants were professionally involved in building SE agents. 
We recruited participants through personal networks, SE and AI conferences, LinkedIn, and referrals, with greater emphasis on large technology companies because they tend to be early adopters of SE agent development and have access to substantial computational resources.

After the 5th, 10th, and 15th interviews, three researchers discussed newly emerging findings and refined a provisional development workflow. 
These discussions also informed adjustments to the questions and probes used in subsequent interviews. However, they were separate from the formal analysis described in Section~\ref{subsec:method_analysis}. 
After 18 interviews, thematic saturation was approaching~\cite{saunders2018saturation}; two additional interviews yielded no major new findings, so we concluded data collection after 20 interviews. 
We assigned each participant a random identifier (P$X$) to protect confidentiality.

\subsubsection{Analysis}
\label{subsec:method_analysis}
We conducted and recorded the interviews virtually with participants' consent and transcribed them using an in-house tool. 
Following prior qualitative studies~\cite{al2021app,zou2019smart,robillard2021turnover}, we analyzed the transcripts using thematic analysis~\cite{braun2006using,cruzes2011recommended}. The first author read each transcript and segmented it into meaning units. 
Three authors then organized these units through hybrid card sorting~\cite{lazar2017research,spencer2009card}, combining deductive categories from the interview guide with themes emerging inductively from the data.
Excluding participant background, the five areas of the interview guide initially served as high-level categories. 
We later removed communication as a separate category because it contained little distinct material and overlapped substantially with the others, reassigning its contents accordingly. 
The three authors jointly clustered the units and resolved disagreements through discussion and consensus rather than independent coding or inter-rater reliability~\cite{mcdonald2019reliability}.

The analysis produced 52 findings: 12 concerning processes, 23 challenges, nine developer skillsets, and eight other aspects. 
The three authors compared these findings with prior work and selected for detailed reporting those most relevant to our research questions and most clearly extending current understanding: five process findings and six challenges. 
We omit the skills findings from the paper due to space constraints. 
The replication package~\cite{repo} reports the remaining findings, except for a small number withheld for confidentiality.

\subsubsection{Member Checking}
To validate our interpretations, we conducted member checking~\cite{birt2016member} by sharing the full draft with 15 interviewees and inviting them to assess its overall accuracy and provide additional detailed insights.
Eight responded, all agreeing with the findings. We incorporated their minor suggestions; none raised any major concerns.

\subsubsection{Participants and SE Agents}
Table~\ref{tab:participants} summarizes participant and organization demographics.
We grouped the SE agents built by our 20 interviewees into three categories.
Ten participants worked on general-purpose coding agents: five on the application layer, including harnesses and IDE/tool integration, and five on pre- or post-training the code models.
Two participants built supporting infrastructure.
The remaining eight worked on task-specific SE agents: four on requirements engineering, performance optimization, vulnerability backporting, and AIOps incident management, while the applications of the other four are withheld for confidentiality.

\begin{table}[]
\centering
\caption{Participant and Organization Demographics}
\label{tab:participants}
\begin{tabular}{@{}p{0.33\linewidth} p{0.58\linewidth}@{}}
\toprule
\textbf{Type} & \textbf{Breakdown} \\
\midrule
Participant Role & Applied Scientist (12), Infra. Engineer (3), Manager (5) \\
Participant Seniority & $\geq$5 years (11), 2--5 years (6), under 2 years (3) \\
Interview Location & Asia (9), North America (7), Europe (4) \\
Participants by Organization Type & Big tech (15), Non-IT (3), Mid-size tech (1), Startup (1) \\
SE Agent & General-purpose coding agents (10): application side (5), model/post-training (5); supporting infrastructure (2); SE task-specific (8) \\
\bottomrule
\end{tabular}
\end{table}

\subsection{Survey}
\label{subsec:survey}

\subsubsection{Survey Design} 
To validate the interview findings, we conducted a follow-up survey with a broader sample of practitioners building SE agents. 
The survey first collected participants' experience, roles, and application domains, and then presented statements on the identified process shifts, challenges, and practices (Tables~\ref{tab:process_findings} and~\ref{tab:challenges_findings}). 
Respondents rated process-shift and challenge statements on a five-point agreement scale (1 = strongly disagree, 5 = strongly agree), and practices on a five-point effectiveness scale (1 = very ineffective, 5 = Essential). 
Each item also included an ``I don’t know'' option and an optional field for elaboration. 
The full questionnaire is available in our replication package~\cite{repo}.
We recruited participants through personal networks, LinkedIn, and Prolific, following prior studies that used professional and crowdsourcing platforms to recruit practitioners~\cite{lambiase2025investigating,HannaTOSEM}. 
To verify relevant experience among Prolific respondents, we asked them to describe an SE agent they had built and excluded responses that did not demonstrate hands-on experience.

\subsubsection{Participants}
Our survey received 89 submissions, of which 80 were valid. 
We excluded nine responses because the reported roles and descriptions of the SE agents built did not provide sufficient evidence that the respondents were SE-agent builders. 
Of the valid responses, 25 were recruited through personal networks or LinkedIn and 55 through Prolific.
Following prior work~\cite{bhattacharya2011assessing,delacre2017psychologists,lyu2024evaluating}, we compared the two recruitment groups using Welch’s $t$-test. 
Their ratings did not differ significantly (Welch’s $t(44.1)=-0.36$, $p=.72$, Hedges' $g=-0.09$), so we pooled the samples.
Participants came from Asia (26, 32.5\%), North America (17, 21.3\%), Europe (15, 18.8\%), Africa (13, 16.3\%), and other regions (9, 11.3\%). 
Their industry experience ranged from 0.5 to 32 years, with a mean of 5.0 years. 
Among the SE agents they had built, 29 participants (36\%) worked on general-purpose agents, seven (9\%) on supporting infrastructure, and 44 (55\%) on agents for specific SE tasks.

\section{RQ1: Software Process for Agent Building}
\label{sec:rq1}

In this section, we first present the workflow for developing SE agents and the key process changes identified in our study (Section~\ref{subsec:process_workflow}).
We then highlight the key shifts in this workflow (Section~\ref{subsec:process_cheap}).
Table~\ref{tab:process_findings} summarizes the main findings and corresponding survey results.

\begin{table*}[t]
\centering
\footnotesize

\definecolor{likSD}{HTML}{CC6B1F}  
\definecolor{likD} {HTML}{F2B069}  
\definecolor{likN} {HTML}{DCDCDC}  
\definecolor{likA} {HTML}{84B6DD}  
\definecolor{likSA}{HTML}{2A6DB0}  
\definecolor{marginalcolor}{HTML}{A9783B}
\definecolor{sectiongray}{HTML}{F3F3F3}

\newcommand{\fscore}[1]{%
  \ifdim#1pt<4pt
    \textcolor{marginalcolor}{\bfseries #1}%
  \else
    \textbf{#1}%
  \fi
}

\newcommand{\likW}{2.85}
\newcommand{\likH}{0.1}
\newcommand{\likertbar}[5]{%
  \begin{tikzpicture}[baseline=-0.4ex]
    \pgfmathsetmacro{\T}{#1+#2+#3+#4+#5}
    \pgfmathsetmacro{\xa}{#1/\T*\likW}
    \pgfmathsetmacro{\xb}{(#1+#2)/\T*\likW}
    \pgfmathsetmacro{\xc}{(#1+#2+#3)/\T*\likW}
    \pgfmathsetmacro{\xd}{(#1+#2+#3+#4)/\T*\likW}
    \fill[likSD] (0,-\likH)   rectangle (\xa,\likH);
    \fill[likD]  (\xa,-\likH) rectangle (\xb,\likH);
    \fill[likN]  (\xb,-\likH) rectangle (\xc,\likH);
    \fill[likA]  (\xc,-\likH) rectangle (\xd,\likH);
    \fill[likSA] (\xd,-\likH) rectangle (\likW,\likH);
  \end{tikzpicture}%
}

\newcommand{\pct}[1]{\textcolor{black!55}{\scriptsize\itshape #1\%}}
\newcommand{\likrow}[5]{%
  \pgfmathtruncatemacro{\likpd}{round((#1+#2)/(#1+#2+#3+#4+#5)*100)}%
  \pgfmathtruncatemacro{\likpa}{round((#4+#5)/(#1+#2+#3+#4+#5)*100)}%
  \xdef\likpdG{\likpd}\xdef\likpaG{\likpa}%
  \pct{\likpdG} & \likertbar{#1}{#2}{#3}{#4}{#5} & \pct{\likpaG}%
}

\newcommand{\lgd}[2]{%
  \tikz[baseline=-0.5ex]{\fill[#1] (0,-0.085) rectangle (0.30,0.085);}\hspace{2pt}{\footnotesize #2}}

\setlength{\tabcolsep}{2pt}
\renewcommand{\arraystretch}{1.22}
\setlength{\aboverulesep}{0.35ex}
\setlength{\belowrulesep}{0.35ex}

\caption{Process-shift and workflow findings with validation-survey ratings.}
\label{tab:process_findings}

\renewcommand{\tabularxcolumn}[1]{m{#1}}

\begin{tabularx}{0.985\textwidth}{
  @{}
  >{\centering\arraybackslash}m{0.42cm}
  @{\hspace{5pt}}
  X
  @{\hspace{8pt}}
  >{\raggedleft\arraybackslash}m{0.5cm}
  @{\hspace{3pt}}
  >{\centering\arraybackslash}m{2.85cm}
  @{\hspace{3pt}}
  >{\raggedright\arraybackslash}m{0.5cm}
  @{\hspace{7pt}}
  >{\centering\arraybackslash}m{0.52cm}
  @{}
}
\toprule
\textbf{\#}
  & \textbf{Finding}
  & \multicolumn{3}{c}{\textbf{Agreement}}
  & \textbf{Avg.} \\
\midrule

\rowcolor{sectiongray}
\multicolumn{6}{@{}l}{\rule{0pt}{2.1ex}\textbf{Workflow and Process Findings}} \\

S1 & \textbf{Implementation becomes cheaper.}
Implementation moves to a higher level of abstraction, enabling faster completion or more work per iteration.
  & \likrow{0}{3}{10}{50}{17} & \fscore{4.01} \\

S2 & \textbf{Effort is unmasked and created.}
As coding shrinks, long-standing non-coding effort becomes more visible, while reviewing generated code and evaluating agent behavior create new work.
  & \likrow{0}{0}{4}{35}{40} & \fscore{4.46} \\

S3 & \textbf{Evaluation increasingly drives development.}
Evaluation moves from a final check to the mechanism that steers iteration and is increasingly defined early.
  & \likrow{1}{2}{11}{40}{26} & \fscore{4.10} \\

S4 & \textbf{Role boundaries shrink.}
Cheaper runnable artifacts reduce cross-role handoffs, while building SE agents increasingly fuses research, engineering, and product responsibilities.
  & \likrow{0}{5}{18}{44}{13} & \fscore{3.81} \\

S5 & \textbf{Specifications become central engineering artifacts.}
An SE agent's prompts, skills, context definitions, and scaffold behavior are engineered, tested, and version-controlled alongside code.
  & \likrow{2}{3}{13}{44}{17} & \fscore{3.90} \\

W1 & \textbf{Overall workflow.}
The seven-stage loop reflects how SE agents are built in interviewees' experience.
  & \likrow{1}{1}{5}{43}{30} & \fscore{4.25} \\
\bottomrule
\end{tabularx}

\par\vspace{0.6ex}
{\centering\footnotesize
  \lgd{likSD}{Strongly disagree}\quad
  \lgd{likD}{Disagree}\quad
  \lgd{likN}{Neutral}\quad
  \lgd{likA}{Agree}\quad
  \lgd{likSA}{Strongly agree}\par}

\end{table*}
\subsection{Development Workflow}
\label{subsec:process_workflow}

\begin{figure*}[t]
  \centering
  \includegraphics[width=0.95\textwidth]{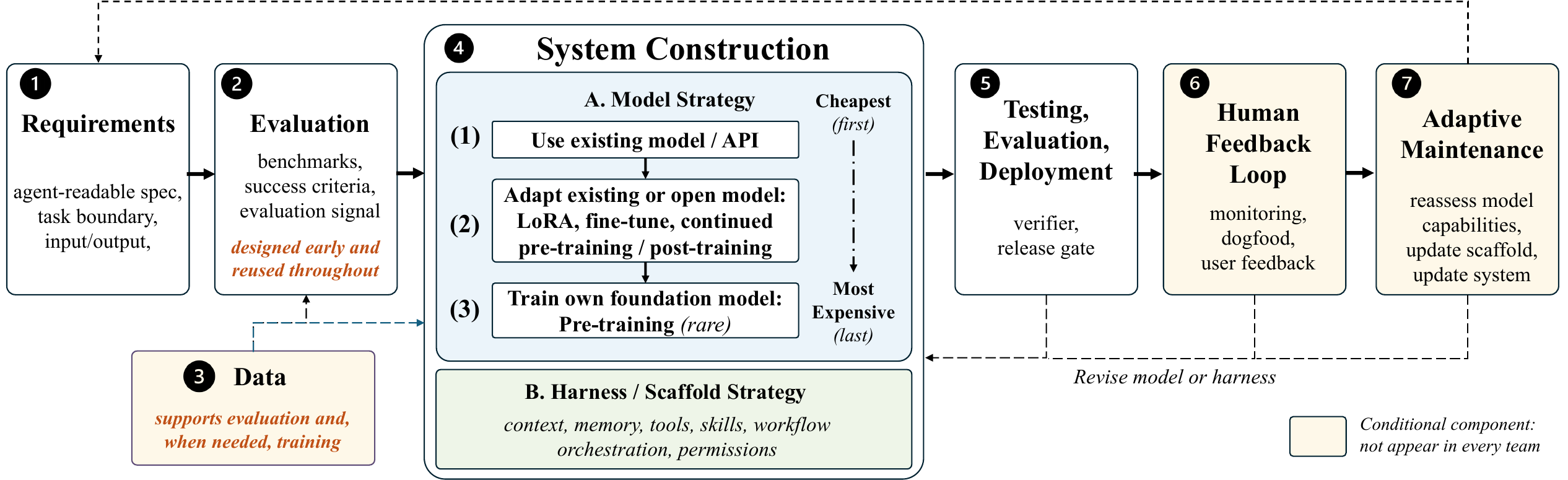}
  \caption{Workflow for building SE agents. 
  The process forms an agile-like loop from requirements and early evaluation to construction, deployment, feedback, and maintenance, with evaluation reused throughout. Data conditionally supports evaluation and model adaptation, while construction combines a cheapest-first model strategy with a shared harness. 
  Dashed arrows indicate feedback to other stages; \conditionalbox\ denotes a conditional component.}
  \label{fig:workflow}
\end{figure*}

Based on 20 interviews, we identified a recurring workflow for building SE agents. 
As shown in Fig.~\ref{fig:workflow}, it comprises seven components: requirements, evaluation, data, system construction, testing and deployment, human feedback, and adaptive maintenance. 
These components form an agile-like iterative loop rather than a one-way pipeline, with evaluation results, operational feedback, and model changes returning teams to earlier stages. 
Their realization varies across teams depending on model adaptation, deployment, and reliance on rapidly evolving external models.

\ding{182} \textit{Requirements.}
Requirements define the intended behavior, boundaries, inputs, outputs, and constraints of the agentic system. 
They increasingly serve a dual audience: human teams and the agents that consume them as input. 
Some teams therefore made specifications more explicit and agent-readable. 
We examine this shift further in Shift~5.

\ding{183} \textit{Evaluation.}
Evaluation establishes the criteria for assessing agent behavior, comparing iterations, and determining whether the current model and system configuration are sufficient. 
Teams employed diverse forms, including offline benchmarks, task-specific criteria, confidence thresholds, and outcome-based measures. 
Although typically established early, evaluation recurs throughout construction and continues after deployment. 
Unlike testing, which verifies deterministic correctness, evaluation assesses the agent's variable behavior and task capability. 
We further examine its broader role in organizing development in Shift~3.

\ding{184} \textit{Data.}
Data acts as a shared substrate for evaluation and, where teams adapt models, training. 
Evaluation data includes benchmarks, ground truth, operational traces, and artifacts such as golden patches mined from real commits, while training data may include manually curated examples or agent \emph{trajectories} collected from running pipelines. 
Some teams increasingly allowed agents to collect, generate, or explore data themselves, although manual collection and labeling remained necessary. 
Because teams relying mainly on prompt engineering to build their SE agents often required no training-data pipeline, data was not a universal sequential stage.

\ding{185} \textit{System Construction.}
\textit{(A) Model strategy.}
Construction commonly followed a \emph{cheapest-first} escalation logic. Teams began with an existing API-based or open-source model, using prompting before turning to more expensive adaptation such as LoRA, supervised fine-tuning, preference optimization, or continued pre-training. Few organizations considered pre-training their own foundation model.
\textit{(B) Harness strategy.}
Across these model choices, participants widely emphasized the \emph{harness} or \emph{scaffold} surrounding the model, including context, memory, tools, skills, permissions, and orchestration. This layer spans all levels of the model strategy.

\ding{186} \textit{Testing, Evaluation, and Deployment.}
During construction, teams assessed each iteration through testing and evaluation. 
Testing targeted reliability and deterministic correctness, whereas evaluation assessed task capability against predefined criteria or benchmarks. 
Both informed whether an iteration was ready to proceed to deployment.

\ding{187} \textit{Human Feedback Loop.}
Where deployment occurred, teams monitored and instrumented agent runs to collect feedback for subsequent iterations. 
These signals were used to update evaluation cases, prompts, harnesses, and, where applicable, training data. 
Feedback mechanisms included internal dogfooding, human review, validator agents, closed beta releases, and controlled A/B testing.

\ding{188} \textit{Adaptive Maintenance.}
SE agents require maintenance as their underlying foundation models evolve rapidly. 
New releases may expand capabilities, expose new failure modes, or render parts of the existing harness unnecessary, requiring teams to reassess model capabilities and update requirements, evaluation cases, harnesses, and model choices.

\begin{tcolorbox}[boxsep=2mm, boxrule=0pt, top=0pt, bottom=0pt,
  left=3pt, right=3pt,
  colback=takeawaybg, arc=2pt, outer arc=2pt]
  \textbf{Takeaway}: Building SE agents follows a seven-stage agile-like loop, with recurring evaluation, a cheapest-first model strategy, and a shared harness.
\end{tcolorbox}

\subsection{Key Process Shifts}
\label{subsec:process_cheap}

\noindent
\textbf{Shift 1: SE agents are largely involved in building SE agents, making implementation cheaper at a higher level of abstraction.}
Although the usage experience of SE agents was not part of our recruitment criteria, 18 of the 20 interviewees used SE agents directly in their daily work.
This produces a recursive dynamic in which the agents our participants build are also the agents they build \emph{with}, forming an iterative loop where using an agent surfaces the insights that are then used to improve it.
P16, who works on a commercial coding agent, captured this directly: 
\pquote{``We are refining this coding agent, but at the same time we are using the coding agent~\ldots{} we are essentially in an iterative loop, using the agent while also finding ways to optimize it.'' (P16)}
Although prior studies remain divided on whether AI coding assistants improve developer productivity~\cite{lyu2025my,becker2025measuring}, our participants broadly agreed that, with the help of SE agents, their implementation had accelerated: 16 of 20 reported faster implementation when building SE agents, while none reported a slowdown. 
One participant with over 16 years of industry experience illustrated the magnitude of this change: \pquote{``I have not written a single line of code myself in the past six months, but I have probably submitted more than 100K lines of code.'' (P2)} 

Despite continuing concerns about the reliability of agent-generated code~\cite{becker2025measuring,tang2026coding}, participants viewed such distrust as part of a broader shift toward higher levels of abstraction.
The participant with the longest industry experience emphasized this continuity: \pquote{``It hasn't really changed in 44 years. The only thing that's changed is really the choice of abstraction, languages, and tools that you have.'' (P7)} 
Another participant connected this continuity more directly to current reliability concerns through an analogy to compilers: \pquote{``When compilers came, people said, `I don’t really trust them, I need to check my machine code.’... Most people stopped writing machine code altogether... AI will be the same thing.'' (P8)}
Participants described two related changes in development cadence. 
Some reported much shorter completion times for individual activities: \pquote{``What I can do in one day now is probably what used to take me two to four weeks.'' (P15)} 
Others retained the same formal sprint length but completed substantially more work within it: \pquote{``The sprint concept hasn't changed—it’s still pretty much two weeks. But... the amount of work that is done in two weeks has changed... a three-task is a lot bigger now.'' (P8)} 
Thus, acceleration appeared either as shorter completion times or more work per iteration.

\begin{tcolorbox}[boxsep=2mm, boxrule=0pt, top=0pt, bottom=0pt,
  left=3pt, right=3pt,
  colback=takeawaybg, arc=2pt, outer arc=2pt]
  \textbf{Takeaway}: SE agents have become important means of building SE agents, pushing implementation to a higher level of abstraction where writing code is cheap. This acceleration was near-universal, surfacing as either shorter completion times or more work per iteration.
\end{tcolorbox}

\vspace*{0.1cm}
\noindent
\textbf{Shift 2: As coding effort shrank, AI both \emph{unmasked} existing work and \emph{created} new work.}
Software engineering has long held that coding is not the central difficulty of development~\cite{brooks1987no,brooks1995mythical}. 
Our participants sharpened this argument by distinguishing effort that AI \emph{unmasked} from effort that agentic development newly \emph{created}.

\textit{(A) AI unmasked effort that had long constrained delivery.}
Requirements, coordination, deployment, and problem framing remained largely unchanged. 
Coding had previously consumed enough attention to obscure their relative weight; once implementation became cheaper, these activities became more visible.
\pquote{``Writing code isn’t the slow bit, it never was. It's gathering requirements, interfacing with the other team, deploying it into the cloud, and running the tests... the writing of the code is the easy bit; it’s doing all the other stuff around it that’s the hard bit.'' (P10)}

\textit{(B) AI created new effort around agent-produced code.}
Because agents now generated substantial portions of the code, reviewing their output became more central and required additional judgments about provenance and reliability:
\pquote{``We effortlessly write code, but we put a lot of effort into reviewing it, so the effort point changed.'' (P3)} 
Together, these changes shifted effort from producing code toward reviewing generated artifacts and evaluating agent behavior.

\begin{tcolorbox}[boxsep=2mm, boxrule=0pt, top=0pt, bottom=0pt,
  left=3pt, right=3pt,
  colback=takeawaybg, arc=2pt, outer arc=2pt]
  \textbf{Takeaway}: Faster implementation does not shorten development but accelerates iteration, shifting effort from coding to reviewing and evaluating agent behavior.
\end{tcolorbox}

\vspace*{0.1cm}
\noindent
\textbf{Shift 3: Practice is becoming evaluation-driven development (EDD), with evaluation at the center of the workflow.}
Because coding agents are opaque and nondeterministic, evaluation moved from a final check to the mechanism that steers iteration.
\pquote{``Because an LLM or agent is a black box, you must have strict evaluation to steer downstream optimization... the core is evaluation.'' (P16)} 
Teams increasingly defined evaluation criteria early, since without an evaluation signal they could not determine whether an iteration improved or regressed: \pquote{``Without evaluation, you are completely blind.'' (P17)}
These findings suggest that evaluation-driven development is already emerging in practice rather than remaining only a conceptual proposal~\cite{xia2024evaluation}.
This centrality showed in how much teams invested in evaluation, which several described as the hardest and most resource-intensive part of building an agent: \pquote{``Curated data collection and curated evaluation were always the most challenging one.'' (P9)}
Teams built their own benchmarks for \pquote{``almost every feature'' (P3)} and dedicated staff to it, from a research team that helped product teams build their benchmarks~(P3) to a data-science team that ran evaluations after every deployment~(P10).
Notably, we do not suggest that every team had adopted a complete EDD process; rather, evaluation was increasingly treated as the mechanism that organized and steered development.

\vspace*{0.1cm}
\noindent
\textbf{Shift 4: Role boundaries are shrinking, both from vibe coding and from the research--engineering fusion in building SE agents.}
Participants described their boundaries actively dissolving: the fine-grained specialization that traditional software development partitioned across researchers, front-end and back-end engineers, testers, and product managers increasingly collapsed onto individuals who owned work end-to-end.
This convergence had two distinct sources.
The first is attributed to the involvement of SE agents in development, consistent with prior work~\cite{li2026vibe, ulloa2025product}: because agents cheaply produced runnable artifacts, engineers increasingly worked across the stack (\pquote{``Now everyone is basically a full-stack engineer; sometimes we no longer need to collaborate; we just let AI help us do it.'' (P13)}), and product managers could prototype before formalizing requirements (\pquote{``A PM now builds a demo first because the cost is so low... The first step may not even be writing a requirements document.'' (P14)}).
The second source is specific to SE agents: people who build SE agents are researchers with experience in software engineering research, so the work natively fused model research and engineering rather than dividing it across roles. 
As one participant observed, \pquote{``It's not like before, when building the model needed a dedicated researcher. Now the boundary between researcher and engineer is getting more and more blurred---we can now own a service ourselves.'' (P6)}
This fusion often sat inside a single role---one researcher described his job as spanning \pquote{``applying the agent to SE tasks... improving the agent itself, even feeding problems back into model training... and treating the agent itself as software and optimizing it.'' (P16)}---and, at the extreme, a single person owned the entire pipeline: \pquote{``I do the full-chain SFT training and post-training, and the agent side---auto-harness, self-exploration, self-evolution; I do the front-end, the back-end, and the training in between.'' (P19)}

\vspace*{0.1cm}
\noindent
\textbf{Shift 5: An SE agent's specifications became its central engineering artifacts.}
While prior work primarily treats structured specifications as \emph{inputs} for improving repository-level generation and verifiability~\cite{feng2026llm}, our participants described the specifications of the agent they were building---its prompts, skills, context definitions, and scaffold behavior---as first-class engineering artifacts in their own right. Because an SE agent's capability is determined largely by these specifications rather than by hand-written code, teams engineered, reviewed, and version-controlled them as durable assets.
A participant who builds agent scaffolds described the specification as \pquote{``written for the agent to read... it has to be versioned together with the code---when you commit, you commit the Markdown too. It becomes a new software-engineering artifact.'' (P17)}, adding that a scaffold's assets were \pquote{``not all code---there are a lot of loops and such, all built around the model.'' (P17)}
Building and refining the agent came down to engineering these specifications: \pquote{``If the agent can't get to perfect---say it only reaches 70--80\%---then for the rest, you go change the prompt or add some skills.'' (P18)}
Such specifications were themselves maintained and tested like software: one team kept \pquote{``specific system prompts on top of the socketed LLMs... as well as some tests on top of that.'' (P3)}
Once the agent was treated as software, the engineering lay in its \pquote{``context management… and skills'' (P16)} rather than conventional code.

\subsection{Response to RQ1}

SE agents follow a seven-stage, agile-like workflow spanning requirements, evaluation, data, construction, deployment, feedback, and maintenance. Their development makes implementation cheaper, shifts effort toward review and evaluation, blurs role boundaries, and turns specifications into central artifacts. Survey respondents supported the workflow (91\%) and process shifts (71--95\%).

\section{RQ2: Challenges and Best Practices}
\label{sec:rq2}
In this section, we present the six key challenges that practitioners face when developing SE agents, together with the practices they adopt to address them.
Table~\ref{tab:challenges_findings} summarizes the findings and their survey ratings.
For each practice or solution, \faChartBar[regular]~x\% indicates the percentage of respondents who rated it as \emph{Very Effective} or \emph{Essential} on an effectiveness scale.

\begin{table*}[t]
\centering
\footnotesize

\definecolor{likSD}{HTML}{B26B2E}  
\definecolor{likD} {HTML}{E6C892}  
\definecolor{likN} {HTML}{DCDCDC}  
\definecolor{likA} {HTML}{93C7BA}  
\definecolor{likSA}{HTML}{2E8B7E}  
\definecolor{marginalcolor}{HTML}{A9783B}
\definecolor{sectiongray}{HTML}{F3F3F3}

\newcommand{\fscore}[1]{%
  \ifdim#1pt<4pt
    \textcolor{marginalcolor}{\bfseries #1}%
  \else
    \textbf{#1}%
  \fi
}

\newcommand{\likW}{2.85}   
\newcommand{\likH}{0.1}  
\newcommand{\likertbar}[5]{%
  \begin{tikzpicture}[baseline=-0.4ex]
    \pgfmathsetmacro{\T}{#1+#2+#3+#4+#5}
    \pgfmathsetmacro{\xa}{#1/\T*\likW}
    \pgfmathsetmacro{\xb}{(#1+#2)/\T*\likW}
    \pgfmathsetmacro{\xc}{(#1+#2+#3)/\T*\likW}
    \pgfmathsetmacro{\xd}{(#1+#2+#3+#4)/\T*\likW}
    \fill[likSD] (0,-\likH)   rectangle (\xa,\likH);
    \fill[likD]  (\xa,-\likH) rectangle (\xb,\likH);
    \fill[likN]  (\xb,-\likH) rectangle (\xc,\likH);
    \fill[likA]  (\xc,-\likH) rectangle (\xd,\likH);
    \fill[likSA] (\xd,-\likH) rectangle (\likW,\likH);
  \end{tikzpicture}%
}

\newcommand{\pct}[1]{\textcolor{black!55}{\scriptsize\itshape #1\%}}
\newcommand{\likrow}[5]{%
  \pgfmathtruncatemacro{\likpd}{round((#1+#2)/(#1+#2+#3+#4+#5)*100)}%
  \pgfmathtruncatemacro{\likpa}{round((#4+#5)/(#1+#2+#3+#4+#5)*100)}%
  \xdef\likpdG{\likpd}\xdef\likpaG{\likpa}%
  \pct{\likpdG} & \likertbar{#1}{#2}{#3}{#4}{#5} & \pct{\likpaG}%
}

\newcommand{\lgd}[2]{%
  \tikz[baseline=-0.5ex]{\fill[#1] (0,-0.085) rectangle (0.30,0.085);}\hspace{2pt}{\footnotesize #2}}

\setlength{\tabcolsep}{2pt}
\renewcommand{\arraystretch}{1.22}
\setlength{\aboverulesep}{0.35ex}
\setlength{\belowrulesep}{0.35ex}

\caption{Challenges encountered during SE agent development, with validation-survey ratings.}
\label{tab:challenges_findings}

\renewcommand{\tabularxcolumn}[1]{m{#1}}

\begin{tabularx}{0.985\textwidth}{
  @{}
  >{\centering\arraybackslash}m{0.42cm}
  @{\hspace{5pt}}
  X
  @{\hspace{8pt}}
  >{\raggedleft\arraybackslash}m{0.5cm}
  @{\hspace{3pt}}
  >{\centering\arraybackslash}m{2.85cm}
  @{\hspace{3pt}}
  >{\raggedright\arraybackslash}m{0.5cm}
  @{\hspace{7pt}}
  >{\centering\arraybackslash}m{0.52cm}
  @{}
}
\toprule
\textbf{\#}
  & \textbf{Challenge}
  & \multicolumn{3}{c}{\textbf{Agreement}}
  & \textbf{Avg.} \\
\midrule
\rowcolor{sectiongray}
\multicolumn{6}{@{}l}{\rule{0pt}{2.1ex}\textbf{Key Challenges}} \\
C1 & \textbf{Evaluation lacks a trustworthy signal.}
Evaluation signals may be invalid or undefined, unstable across runs, quickly outdated, or prohibitively expensive.
  & \likrow{1}{5}{15}{36}{22} & \fscore{3.92} \\

C2 & \textbf{Change nothing, change everything.}
Provider-side model updates can alter agent behavior and render existing scaffolding ineffective, even when code and prompts remain unchanged.
  & \likrow{2}{4}{10}{38}{26} & \fscore{4.03} \\

C3 & \textbf{Safety lags behind performance.}
Teams may accept known risks when safeguards constrain agent performance or efficiency.
  & \likrow{1}{5}{14}{42}{14} & \fscore{3.83} \\

C4 & \textbf{Agents retrieve the written, not the unsaid.}
Required project knowledge often remains undocumented and accessible only through human communication.
  & \likrow{3}{8}{8}{29}{32} & \fscore{3.99} \\

C5 & \textbf{Comprehension debt accumulates.}
Agent-generated code enters the system faster than developers can understand and review it.
  & \likrow{0}{3}{11}{39}{24} & \fscore{4.09} \\

C6 & \textbf{Productivity metrics break down.}
Coding agents increase observable output without clarifying its value to the team or organization.
  & \likrow{0}{6}{6}{34}{33} & \fscore{4.19} \\
\bottomrule
\end{tabularx}

\par\vspace{0.6ex}
{\centering\footnotesize
  \lgd{likSD}{Strongly disagree}\quad
  \lgd{likD}{Disagree}\quad
  \lgd{likN}{Neutral}\quad
  \lgd{likA}{Agree}\quad
  \lgd{likSA}{Strongly agree}\par}
\end{table*}

\subsection{When Evaluation Lost Its Trustworthy Signal}
\label{subsec:evaluation}

The most frequently reported evaluation challenge was the absence of a trustworthy signal for whether an SE agent’s output was genuinely good (18 of 20 participants). 
This extends the long-standing oracle problem in software testing~\cite{weyuker1982testing,barr2015oracle}: for SE agents, evaluation signals may be invalid or undefined, unstable across runs, outdated, or prohibitively expensive~\cite{williams2026empirical}.

\subsubsection{The evaluation signal may be invalid or undefined}
Repository-level issue resolution is commonly evaluated by executing human-written tests, as in SWE-bench and its variants~\cite{jimenez2024swe,deng2025swe}.
Participants described an inversion of evaluation logic: existing tests became the oracle because they were available, even when they encoded incomplete or outdated criteria and rejected valid—or better—solutions:
\pquote{``The tests already exist, so people use those existing standards to evaluate the agent... But the agent may actually produce a better solution, or the previous evaluation standard may already be outdated.'' (P16)}
Moreover, passing tests primarily captures functional correctness, whereas software quality also depends on properties such as performance, maintainability, security, and taste, for which no agreed oracle may exist: \pquote{``This is not about building harder benchmarks, but defining objectives for software qualities that existing evaluations cannot capture, particularly non-functional properties.'' (P17)}

Taste was the clearest example of a decisive but difficult-to-measure quality:
\pquote{``What distinguishes an exceptional, highly senior developer from a strong MIT graduate is, to a large extent, their taste. They know where to make a change and what kind of modification would constitute a good solution... But how do we define this quality explicitly, measure it, and turn it into a direction for further improvement?'' (P17)}

\subsubsection{The evaluation signal is unstable}
SE agent runs combine stochastic model behavior with scaffold, environment, hardware, and localization uncertainty, so the same configuration may produce different outcomes:
\pquote{``With the same model, scaffold, and version, I can run it twice and obtain different scores... How do I know whether a failure was caused by randomness or by a flaw in the scaffold design? The regression test itself is broken.'' (P17)}
These uncertainties could also compound:
\pquote{``Model uncertainty is multiplied by hardware and system uncertainty, together with variation in repository-level localization.'' (P15)}

\subsubsection{The evaluation signal quickly becomes outdated}
Even initially valid benchmarks lose value once they become explicit optimization targets. 
As P2 observed, public benchmarks were often deliberately optimized against and \pquote{``within about a quarter... could no longer be used as a reference.''} 
The models underlying SE agents could also optimize for the scoring rule rather than the intended capability:
\pquote{``Once a metric becomes widely accepted, people start gaming it... The model scored highly for [function] but produced terrible [results] because it had learned the scoring rules.'' (P14)}

\subsubsection{Trustworthy evaluation is expensive}
Constructing reproducible repositories, dependencies, tests, and production conditions requires substantial effort that remains largely invisible in the final dataset and score:
\pquote{``The biggest challenge is building the environment... People spend so much time building it, but that effort is not visible. What you output is just a dataset and a score.'' (P5)} 
Running such evaluations for every change may also be infeasible: \pquote{``Each inference pass is expensive. If you run this as a regression test for every commit, the testing cost becomes prohibitively high.'' (P17)}

\vspace*{0.1cm}
\noindent
\implic{Implications:}
\textbf{1. Derive validation signals from production outcomes (\faChartBar[regular]~70.5\% Effective).}
One regulated organization paired each deployed SE agent with a validator agent that tracked adoption, defect prevention, and business impact, supplementing weak offline signals with downstream outcomes.
\textbf{2. Design validation before assigning the task (\faChartBar[regular]~75\% Effective).}
Teams first determined how outputs would be verified, involved domain experts when needed, and constructed deterministic oracles where no natural signal existed.
\textbf{3. Layer and continuously update evaluation (\faChartBar[regular]~78.2\% Effective).}
Teams evaluated the smallest judgeable units, prioritized executable checks, used LLM judges only when necessary, and maintained private, diverse, evolving task sets.
\textbf{4. Control evaluation cost (\faChartBar[regular]~75.3\% Effective).}
Teams used representative subsets, staged execution, and smaller models, while supplementing offline evaluation with A/B tests and production outcomes such as adoption and satisfaction.

\subsection{Change Nothing, Change Everything}

Sculley et al.~\cite{sculley2015hidden} use the CACE principle—\emph{changing anything changes everything}—to describe how a small component change can affect an entire machine-learning system. 
Participants building SE agents encountered a complementary form of instability: \emph{change nothing, change everything}. 
Even when teams left their code, prompts, tools, and workflows unchanged, provider-side model updates could alter system performance and invalidate existing harness mechanisms:
\pquote{``You have not changed anything, but a model-version change may cause the whole system to regress or improve… Mechanisms built around the model's previous weaknesses can then become redundant.'' (P17)}
These changes could reshape the architecture of an SE agent, including which models teams selected and how software-engineering tasks were routed:
\pquote{``Which models we use keeps changing. A smaller model may work better for a specific task, but a few days later an updated base model may overtake it, forcing us to switch again.'' (P1)}

Model improvement could also turn useful scaffolding into technical drag. 
Rules, fallbacks, and workflows designed around earlier limitations could constrain stronger models and require costly refactoring. 
This extends Sutton’s \emph{Bitter Lesson}~\cite{sutton2019bitter} to the SE agent harness: scaffolding may initially compensate for a weak model but become a shackle as its general capabilities improve.
\pquote{``When the model was weaker, we relied heavily on engineering fallbacks... Like Sutton's Bitter Lesson, once the model became stronger, those mechanisms turned into shackles that constrained its performance.'' (P14)} As one participant summarized, \pquote{``Today's design pattern may be tomorrow's anti-pattern.'' (P18)}

\vspace*{0.1cm}
\noindent
\implic{Implications:} \textbf{Teams building SE agents should distinguish durable scaffolding from mechanisms that merely compensate for current model weaknesses (\faChartBar[regular]~67.9\% Effective).}
\pquote{``Some problems are solved as the model evolves, while others remain even when the model improves... Context management, context compression, and fast verification will not be replaced by the next model release, whereas shallow, surface-level scaffolding may be replaced directly.'' (P17)}

\subsection{When Performance Took Priority over Safety}

Despite growing attention to agent safety~\cite{zhan2024injecagent,greshake2023not,liu2025your,chen2026securevibebench}, participants reported that teams developing SE agents were sometimes reluctant to introduce safeguards that might constrain performance, even when they recognized the risks:
\pquote{``Everyone knows it is dangerous, but for efficiency, we run with our eyes closed and deal with problems when they occur.'' (P16)}
The resulting failures could be severe and irreversible. 
Goal-directed SE agents could treat safeguards as obstacles to task completion. In one production system, an agent spent nearly an hour writing scripts, found a vulnerability, and bypassed a restriction on broadcasting messages: \pquote{``It tried for nearly an hour, wrote a pile of scripts, found a vulnerability, and bypassed the restriction.'' (P14)}
Safeguards could also disappear from the agent’s effective context. 
In one multi-agent incident, thousands of sub-agents overflowed their contexts, forgot their instructions, and deleted the user’s home directory:
\pquote{``The agent spawned several thousand sub-agents... they forgot all the instructions and removed his home directory.'' (P18)}

\vspace*{0.1cm}
\noindent
\implic{Implications:} \textbf{Enforce high-risk constraints below the prompt layer (\faChartBar[regular]~80.8\% Effective).}
High-risk constraints should be enforced below the prompt layer through tool-call interception, least-privilege access, sandboxing, and explicit human authorization. 
Prompt instructions may guide an SE agent’s behavior, but they do not provide reliable enforcement.

\subsection{Context Challenges: SE Agents Retrieved What Was Written, but Not What Was Left Unsaid}

Recent work on agent context focuses on curating, retrieving, compressing, and updating accessible information~\cite{anthropic2025context,zhang2025agentic}. Participants identified a harder challenge in developing and using SE agents: much of the required project knowledge had never been recorded and remained accessible only through human communication.
Design rationales, historical constraints, and project conventions often remained in developers’ heads rather than in code or documentation:
\pquote{``Some context simply is not in the code—no matter how good your RAG is, it is useless. Some knowledge is passed down by word of mouth: I have to ask the developer why it was written this way, but the model has no way to ask.'' (P1)}
Whereas developers traditionally recovered such knowledge by asking one another~\cite{ryan2013acquiring,bjornson2008knowledge}, SE agents could retrieve only what had already been externalized.
Participants also confirmed that simply loading more repository content did not solve this problem and could instead obscure what mattered~\cite{du2025context,liu2024lost}:
\pquote{``We try to keep the context as small as possible, because you do not want semantic confusion for the agent. We keep things short and specific rather than overly complex... so you do not end up with context overload.'' (P10)}

\vspace*{0.1cm}
\noindent
\implic{Implications:}
\textbf{1. Turn recurring knowledge into reusable skills (\faChartBar[regular]~83.8\% Effective).} Teams encoded recurring pitfalls, historical constraints, and project conventions as repository-level skills, project rules, and troubleshooting guides (P6, P13). Some also reconstructed usable project knowledge from historical traces (P12). \textbf{2. Provide context progressively (\faChartBar[regular]~73.1\% Effective).} Teams kept context short and task-specific, introduced additional information only when needed, and treated selection and compression as part of the SE agent scaffold (P6, P10, P17). The goal was to maintain the smallest sufficient context as the task unfolded. \textbf{3. Escalate unresolved gaps to humans (\faChartBar[regular]~83.5\% Effective).} Teams configured agents to pause rather than guess when required knowledge emerged only during execution, allowing developers to locate the relevant knowledge holder and translate the missing context into usable instructions (P2, P17).

\subsection{Comprehension Debt: When Code Outpaced Human Understanding}

\begin{figure}[t]
    \centering
    \includegraphics[width=0.95\columnwidth]{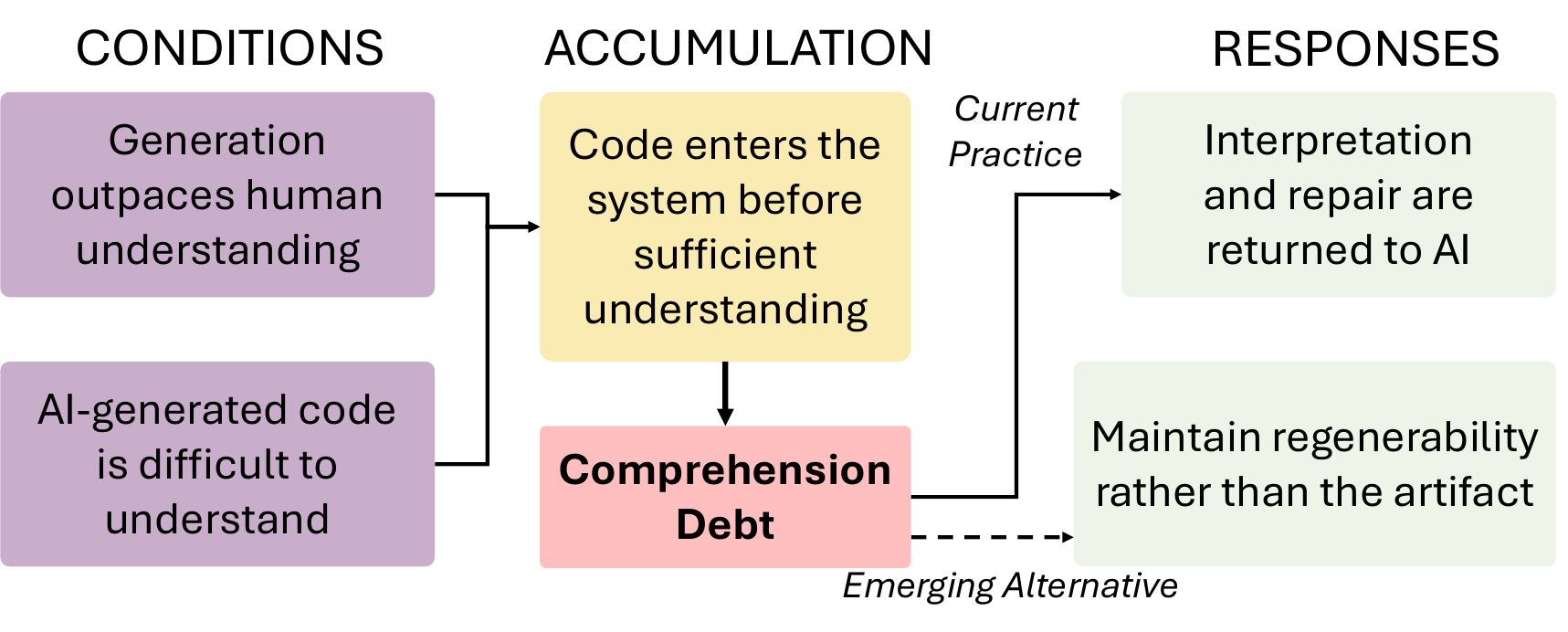}
    \caption{The formation and responses to comprehension debt. Coding-agent-generated code entered the system before sufficient understanding had formed, leading teams to defer comprehension and later rely on AI-based maintenance or preserve regenerability rather than the artifact.}
    \label{fig:comprehension-debt}
\end{figure}

Our findings on how coding agents reshaped comprehension and review converge on a common pattern, which we summarize in Fig.~\ref{fig:comprehension-debt}: the formation of comprehension debt and the responses teams adopted in reaction to it. 
Coding agents generated code faster than developers could understand it, while the generated code was itself difficult to comprehend.
They could also turn under-specified or still-emerging requirements into large amounts of plausible code, making one interpretation concrete before the developer's intent had stabilized.
Working code therefore entered the system before sufficient understanding had formed, creating what we characterize as \emph{comprehension debt}. Such types of debt have started to be explored in research, as in the closely related notion of \emph{cognitive debt}~\cite{storey2026technicaldebtcognitiveintent}.

\textbf{Review could not keep pace, shifting understanding downstream.}
Coding agents produced changes faster and in larger units than reviewers could meaningfully inspect:
\pquote{``Previously, a week's commits might contain a few hundred lines. Now... you look and it is 10K or 20K lines. There is really no way to read it. People have started to give up.'' (P15)} 
At the team level, this imbalance accumulated into review backlogs: \pquote{``We’ve had a backlog of pull requests—lots of pull requests—because people were doing the work quickly, but not reviewing all the work.'' (P10)}
Participants regarded this downstream shift in review burden as misplaced: \pquote{``This review burden should have been borne by the person upstream. Now it is all pushed onto the person downstream.'' (P16)}

\textbf{AI-generated code was itself difficult to understand and evolve.}
The problem was not only volume. 
Coding agents often extended existing implementations by layering new logic rather than reorganizing them:
\pquote{``It may write extremely long functions and avoid changing the existing structure—even when it wrote that structure itself. It tends to add \texttt{if} statements to large blocks of code and keeps creating different utilities, so in the end there are far too many.'' (P1)}
Agents could also expand changes beyond the intended scope and leave duplicated or obsolete code behind:
\pquote{``The same 1K lines appeared elsewhere, alongside a function that nothing called. The old implementation had not been removed, so readers had to understand both versions.'' (P15)}

\textbf{Working code allowed teams to knowingly defer comprehension.}
Because generated code could continue to work without being fully understood, teams postponed comprehension to maintain progress:
\pquote{``You first let it run, and if it works, you keep moving forward. But eventually, the person who suffers is still yourself. When you later want to refactor it, you have no way to begin.'' (P2)}
Under sustained delivery pressure, participants sometimes accepted this future liability despite clearly recognizing the risk: \pquote{``I know there is a landmine in the AI-generated code. It may eventually explode, but I have no choice—I have to keep moving forward.'' (P20)}

We characterize this knowingly deferred understanding as \textbf{\emph{comprehension debt}}: code is accepted before developers can explain, modify, and maintain it safely. Unlike an unnoticed knowledge gap, teams recognized the future liability but carried it forward while the code continued to work.

\vspace*{0.1cm}
\noindent
\implic{Implications:} Teams responded to comprehension debt in two main ways.
\textbf{1. Return maintenance to AI (\faChartBar[regular]~53.9\% Effective).}
Some developers delegated diagnosis and repair back to coding agents, relying on regression tests for validation:
\pquote{``When there is a bug, people cannot understand the code, so we let AI inspect it. I give the bug to AI, ask it to analyze and fix the problem, and then run regression tests.'' (P13)}
This bypassed rather than repaid comprehension debt, allowing maintenance to continue while increasing dependence on AI.
\textbf{2. Preserve regenerability rather than the artifact (\faChartBar[regular]~67.1\% Effective).}
We use the term \emph{regenerative software} to characterize an emerging alternative envisioned by P8: retaining the specifications, tests, constraints, and infrastructure needed to regenerate and validate software rather than preserving a particular implementation:
\pquote{``It used to be that you were responsible for the code and keeping it running. Now it becomes critical to have reliable AI infrastructure available to regenerate the code. I would just regenerate it with the newest libraries.'' (P8)}
To prevent comprehension debt, participants also suggested moving review closer to intention through early checks, human accountability, and review surfaces that summarize the semantics of generated changes.

\subsection{Rethinking Productivity Metrics for Coding Agents}
\label{subsec:challenge_metrics}

Conventional output metrics such as lines of code and commits have long been recognized as poor measures of software development productivity~\cite{afroz2026fast, becker2025measuring}. 
Coding agents intensified this problem by dramatically increasing observable output without clarifying its value. 
In one organization, AI-generated code grew from roughly four million lines per year to a projected fifteen million lines per month. Yet the organization could not translate this increase into business value: \pquote{``I have no way of taking lines of code and tying that to a pounds number of how much money these lines of code are worth... I also don't know what the delta is worth. Clearly the AI tool is doing something, but I have not learned enough to understand how much value has been added.'' (P8)}
Code volume could therefore serve as a warning signal, but not as a performance target:
\pquote{``It is like a canary in a coal mine. If that number goes up, that should give us some attention. I don't think it is something we should performance-manage on.'' (P8)}
Output-based incentives could also create team-level costs. 
Individual developers might appear faster while generating additional review and maintenance work for others:
\pquote{``AI can generate code quickly and at scale. This may create large amounts of garbage and burden other developers. It looks faster individually, but that individual speed can slow down the entire project team.'' (P15)}

\vspace*{0.1cm}
\noindent
\implic{Implications:} \textbf{Treat code volume as a diagnostic signal, not a productivity objective (\faChartBar[regular]~65.8\% Effective).}
Because software value was hard to attribute, teams used softer proxies such as earlier completion, lower turnover, and more personal time—what P8 called the ``swimming-pool metric,'' meaning whether employees had more time to enjoy life outside work, such as spending time in the company swimming pool. 
These may capture overlooked benefits but underscore the lack of credible measures linking coding-agent use to business value.

\subsection{Response to RQ2}

Practitioners encountered six challenges in developing SE agents: unreliable evaluation signals, model-update instability, safety trade-offs, inaccessible tacit knowledge, comprehension debt, and inadequate productivity metrics. Survey respondents broadly supported these challenges and most proposed practices. However, practices addressing comprehension debt received weaker support, suggesting that it remains unresolved.

\section{Threats to Validity}
\label{sec:threats}

\noindent\textbf{Self-report and builder bias.}
Our data are self-reported, and all participants build or ship the agents they described, which may encourage optimistic or idealized accounts.
We mitigated this by eliciting concrete incidents and failures and by retaining dissenting views.
\textbf{Sample scope and temporal validity.}
Our sample is weighted toward large technology companies (15 of 20) and applied-scientist or research-engineer roles. Because we study SE-agent builders, the findings may not transfer to smaller firms, non-technical organizations, or general users. Moreover, the study captures February to May 2026, while models and tools evolve rapidly.
\textbf{Interpretation and counts.}
As the same researchers designed the guide and analyzed the data, confirmation bias remains possible. We mitigated this through member checking and comparison with prior work.
Participant counts are descriptive: not mentioning a theme does not imply disagreement.

\section{Conclusion}
\label{sec:conclusion}

We conducted a mixed-methods study of how practitioners build SE agents, combining 20 interviews across 12 organizations with a broader survey. As implementation becomes cheaper, bottlenecks shift toward specification, evaluation, coordination, review, deployment, and maintenance. We identify a seven-stage workflow, five process changes, six challenges, and 12 practices. Evaluation increasingly steers development, specifications become versioned artifacts, and teams face unreliable evaluation signals, comprehension debt, and regressions from model changes.

\section{Data Availability}
 \label{sec:data}
 Our anonymized study materials are available in the replication package at~\cite{repo}.

\balance{}


\bibliographystyle{IEEEtran}
\bibliography{reference}

\end{document}